\documentclass[prb,nofootinbib,superscriptaddress,twocolumn,showpacs]{revtex4-1}
 
\usepackage{bm}
\usepackage{amsfonts}
\usepackage{amssymb}
\usepackage{amsmath}
\usepackage{amsthm}
\usepackage{hyperref}
\usepackage{times}
\usepackage{caption}
\usepackage{subcaption}

\usepackage[pdftex]{color,graphicx}

\newcommand{\up}{\uparrow}
\newcommand{\down}{\downarrow}

\let\oldmarginpar\marginpar
\renewcommand\marginpar[1]{\-\oldmarginpar[\raggedleft\tiny\color{red} #1]%
{\raggedright\tiny #1}}

\graphicspath{{Figures/}}

\begin{document}

\title{Detecting Majoranas in 1D wires by charge sensing}
	
\author{Gilad Ben-Shach}
\affiliation{Department of Physics, Harvard University, Cambridge, MA 02138}
\author{Arbel Haim}
\affiliation{Department of Condensed Matter Physics, Weizmann Institute of Science, Rehovot 76100, Israel}

\author{Ian Appelbaum}
\affiliation{Center for Nanophysics and Advanced Materials, University of Maryland, College Park, Maryland 20742-4111, USA}

\author{Yuval Oreg}
\affiliation{Department of Condensed Matter Physics, Weizmann Institute of Science, Rehovot 76100, Israel}

\author{Amir Yacoby}
\affiliation{Department of Physics, Harvard University, Cambridge, MA 02138}

\author{Bertrand I. Halperin}
\affiliation{Department of Physics, Harvard University, Cambridge, MA 02138}

\date{\today}

\begin{abstract}
The electron number-parity of the ground state of  a semiconductor narowire proximity-coupled to a bulk superconductor can alternate between the quantised values $\pm 1$ if parameters such as the wire length $L$, the  chemical potential $\mu$ or  the magnetic field $B$ are varied inside  the topological superconductor phase. 
The parity jumps, which may be interpreted as changes in the occupancy of the fermion state formed from the pair of Majorana modes at opposite ends of the wire, are accompanied by jumps $\delta N$ in the charge of the nanowire, whose values decrease exponentially with the wire length. 
We study theoretically the  dependence of $\delta N$ on system parameters, and compare the locations in the $\mu$-$B$ plane  of parity jumps when the nanowire is or is not proximity-coupled to a bulk superconductor.   
We show that, despite the fact that the wave functions of the Majorana modes are localised near the two ends of the wire, the charge-density jumps have spatial distributions that are essentially uniform along the wire length, being proportional to the product of the two Majorana wave functions. 
We explain how  charge measurements, say by an external single-electron transistor,  could reveal these effects.
Whereas existing experimental methods require direct contact to the wire for tunneling measurements, charge sensing avoids this issue and provides an orthogonal measurement to confirm recent experimental developments.
Furthermore, by comparing density of states measurements which show Majorana features at the wire ends with the uniformly-distributed charge measurements, one can rule out alternative explanations for earlier results.
We shed light on a new parameter regime for these wire-superconductor hybrid systems, and propose a related experiment to measure spin density.
\end{abstract}


\maketitle

%
\section{Introduction}

The isolation of zero-energy Majorana  modes is an essential step in various proposals to perform topologically protected quantum computation~[\onlinecite{nayak2008non}].
The existence of localised Majorana modes has been predicted in several condensed matter systems, but definitive detection of such modes remains an open challenge~[\onlinecite{moore1991nonabelions,RiceSigrist,fu2008superconducting,volovik2009universe,SilaevVolovik,PhysRevLett.101.120403,sau2010generic}].

A promising physical system for realizing these modes consists of  a  one-dimensional (1D)  semiconductor wire with strong Rashba spin-orbit coupling, coupled to a bulk  s-wave superconductor (SC), and with a strong applied magnetic field~[\onlinecite{PhysRevLett.105.177002, PhysRevLett.105.077001}]. 
Under appropriate conditions, this system can enter  a ``topological" state,  which would exhibit  isolated Majorana fermions at the wire ends.
The condition for a wire with strong spin-orbit interaction to enter this topological regime, is $E_{\rm Z}^2 > \Delta^2+\mu^2$, where $E_{\rm Z}$ is the Zeeman energy,  proportional to  the applied magnetic field $B$, while $\Delta$ is the induced superconducting pair potential in the wire, and $\mu$ is the chemical potential of the wire, measured  relative to the electron energy at wave vector $k=0$ when $E_{\rm{Z}}=\Delta=0$. 

For an infinitely long wire in the topological regime, the wire has two possible ground states  which are perfectly degenerate.  
The Majorana modes appear at the ends of the wire as zero-energy mid-gap states in the Bogoliubov-deGennes (BdG) spectrum.
Moreover, in this limit the charge density distribution is precisely the same in the two ground states.
For a long but finite wire, the two lowest-energy states of the wire will generally not be perfectly degenerate, but will be split by a small amount, which decreases exponentially as the wire becomes long.  
Similarly, the charge density distributions in the two states will differ by a small amount. 

Since fermion number is conserved mod 2 in the Hamiltonian of the system, the number parity is a good quantum number, which differs in the two competing ground states.  
We can classify the parity by the eigenvalue of the number parity operator, $\pm 1$, and we call these even/odd respectively.
If parameters such as $B$ or $\mu$ or the length $L$ of the system are varied, the energies of the even and odd-parity states can cross, so  the parity of the true ground state can jump discretely between even and odd.

Since the total charge  on the nanowire is not conserved, it is not a good quantum number, and its expectation value, in general, will  not be an integer as the ground state will be a superposition of components with different electron number.
For a finite wire,  there  will be a small but non-zero jump in the total electron number, whenever the parity changes, but the size of the jump can  be much less than one electron charge.  
Between these jumps, the average number of electrons  will vary continuously with the system parameters. 

Although the quantum operators for Majorana modes  do not obey the commutation relations of a normal Dirac fermion creation or annihilation operator, one can construct a  proper annihilation operator from a linear combination of the two Majorana operators at opposite ends of the wire.   
Following a BdG description, the difference between the even and odd parity many-body ground states is equivalent to whether the fermion state corresponding to this annihilation operator is  occupied or not.  
Moreover, the energy  difference between the two ground state energies  is just the BdG energy of this single fermion state.  
Since eigenstates of the BdG equation occur in pairs with energies that differ by a sign,  we may say  that the degenerate zero-energy state is  split in the finite wire, into states of positive and negative energy, due to a small overlap between the Majorana wave functions localised at the two ends.  
Jumps in the parity of the ground state occur when this energy splitting passes through zero. 
The charge difference between the even and odd parity ground states  is equal to the net charge carried by the BdG fermion state, which can be non-zero when the constituent Majorana wave functions overlap.

The purpose of the present paper is to explore in some detail the regions in the phase diagrams where parity jumps are expected, as well as the size of the jumps in electron charge expected at these transitions. 
We also compute the spatial distribution of the jumps in charge density. 
Although the Majorana wave functions, and hence the tunneling density of states, are peaked at the wire ends, we show that the discontinuity in  charge density arising from the overlap of the Majorana wave functions is spread essentially uniformly along the wire.
Changes in total charge and charge density can be measured experimentally using charge sensing techniques.

We note that the number parity of the nanowire  can change when a parameter is varied on laboratory time scales, even though the model Hamiltonian conserves parity, even in the absence of coupling to a normal lead.
This is due to the presence of a small number of thermally activated quasiparticles in the bulk superconductor.
These can be excited across the gap of the SC, or might result form hopping between localised states within the bulk SC.

In an important portion of the topological regime (see Sec.~\ref{sec:SpectrumParity} below), it is predicted that the energy splitting of the Majorana modes will vary as 
 $\delta E \sim \exp(-L/\xi) \cos(k_{\rm F} L)$, where $\xi$ is the induced superconducting  coherence length, $k_{\rm F}$ the Fermi wavevector, and $L$ the length of the wire~[\onlinecite{smokinggun}].  
Theoretically, the easiest way to probe this oscillatory  splitting might be to vary $L$, bringing the ends closer together.
In practice, however, the wire has a fixed length.
It can be effectively shortened in discrete steps by depleting pieces of it using external gates, but local gating may lead to other unforeseen consequences.

Alternatively, an experiment can vary $k_{\rm F}$ to access the oscillations, and $\xi$ to exhibit the exponential envelope.
Both $k_{\rm F}$ and $\xi$ depend on the chemical potential, which can be controlled with a global backgate, and the applied external magnetic field.
It has therefore been suggested in~[\onlinecite{smokinggun}] to look for signatures of this dependence.
We demonstrate  that charge-sensing measurements could reveal such oscillations, and  thus may be a natural next step in the search for experimental verification of the elusive Majorana end modes. 

Many recent experiments~[\onlinecite{Mourik25052012,MotyExpt1,Churchill1,Willy1,deng2012anomalous,Rokhinson2012fractional}] have  probed these one-dimensional semiconductor-superconductor hybrid systems by studying electron transport through the nanowire.
Such transport experiments are very promising, but other physical mechanisms have been offered as explanations for the observed effects~[\onlinecite{pikulin2012zero,Flensberg2010,kells2012low,Liu2012}].
In particular, end effects, including Kondo physics~[\onlinecite{PhysRevLett.109.186802}], can cause  zero-bias peaks similar to the ones observed.
The alternate explanations suggest that the transport measurements may be sensitive to other effects beyond the possible Majorana modes predicted to exist at the ends of the wire.
Furthermore, recent studies suggest that contact with a normal metal lead reduces the induced pair potential in the wire~[\onlinecite{Stanescu}].
An alternative experiment, using a capacitive AC measurement of the density of states, was proposed in~[\onlinecite{Ian}].

Lin \textit{et. al.}~[\onlinecite{PhysRevB.86.224511}] proposed an alternate experiment to probe the Majorana states by charge sensing using a single electron transistor (SET).
As addressed above and assumed in our discussion, such a measurement does not require tunneling to a normal lead, which could avoid some of the complications encountered in previous experiments.
Although the authors of~[\onlinecite{PhysRevB.86.224511}] present numerical calculations that illustrate the charge density jumps associated with Majorana states in various cases, we present here a more detailed analysis of these features.
While we employ a simplified model of the physical system, in which we neglect the Coulomb interactions between electrons in the nanowire, we believe that results presented hold for real systems, and the effect should be visible in a realistic experiment.
We address the effects of interactions in section~\ref{sec:Interactions} below.
Various regimes in parameter space are discussed.
We also address how to extract relevant system parameters using this technique, demonstrating that this experimental technique has other applications beyond the intended goal of detecting split Majorana end states.

We stress that a scanning charge measurement showing the additional charge spread across the wire, combined with a scanning tunneling measurement, can rule out alternative explanations of end effects for the previously observed features of Majorana physics.

The rest of this paper is structured as follows.
In Sec.~\ref{sec:modeling}, we present the model and relevant parameters. 
We then discsuss the spectrum and number parity of systems with and without induced superconductivity in Sec.~\ref{sec:SpectrumParity}.
We address the charge of the wire in three sections, beginning with an analytic analysis of the split Majorana modes in Sec.~\ref{sec:Jumps}, followed by a numerical calculation of the total change in charge in Sec.~\ref{sec:TotalCharge}, and then a discussion of the spatial distribution of the charge along the wire in Sec.~\ref{sec:ChargeDistribution}.
We end with an analysis of jumps in spin density in Sec.~\ref{sec:Spin}, a discussion of the effects of electron-electron interactions and screening in Sec.~\ref{sec:Interactions}, and an overview of future experiments in Sec.~\ref{sec:Experiment}.

\section{Modeling}
\label{sec:modeling}
We model the wire using a standard BdG Hamiltonian:
\begin{align}
H_{\text{BdG}}&= \left(-\frac{\partial_x^2}{2m}-\mu (x)\right) \tau_z + E_{\text{Z}}\sigma_z \tau_z + \nonumber \\
&+ i\alpha \partial_x \sigma_y \tau_z + \Delta  \sigma_y \tau_y
\label{eq:Hamiltonian}
\end{align}
where $\alpha$ is the Rashba spin orbit parameter, and $E_{\text{ Z}}=-g\mu_B B/2$ is the Zeeman energy in an applied magnetic field $B$, with g-factor $g$, and $\mu_B$ the Bohr magneton~[\onlinecite{Alicea2012}].
We have chosen the pair potential, $\Delta$, proximity induced from the superconductor, to be positive and real.
If tunneling between the SC and nanowire is strong, $\Delta$ can approach $\Delta_{SC}$, the gap of the bare SC, whereas if the tunneling is weak, $\Delta$ can be arbitrarily small.
The $\tau_j$ and $\sigma_j$ are Pauli matrices in particle-hole and spin space respectively.
%
%
We choose $\mu(x)=\mu$ constant along the length of the wire.

We find the eigenvalues, $\epsilon^{\nu}$, and the corresponding eigenfunctions,
\begin{align}
\psi^{\nu}(x) \equiv \left( u^{\nu}_{\uparrow},u^{\nu}_{\downarrow},v^{\nu}_{\uparrow},v^{\nu}_{\downarrow} \right)^{\text{T}}.
\end{align}

We can then compute the average charge density at each site at finite temperature, with $f(\epsilon)$ the Fermi-Dirac distribution:
\begin{align}
\label{eq:ChargeDensity}
	\left\langle \rho (x) \right\rangle_T  = \sum_{\nu,\sigma} |u_{\sigma}^{\nu}(x)|^2 f(\epsilon^{\nu}) + |v_{\sigma}^{\nu}(x)|^2 f(-\epsilon^{\nu}),
\end{align}  
where the sum is over states with $\epsilon^{\nu}>0$.
We can tune $B$ and $\mu$, and calculate the induced change of the charge.

To apply our model numerically, we rewrite the Hamiltonian on a 1D lattice with total length $2\mu \text{m}$.
For Figs.~1-3, we use 80 sites, and for Fig.~4 we use 160 sites.
Both give a band-width larger than all other energy scales, as desired for numerical accuracy. 
The figures shown in this paper were computed using realistic parameters that might be appropriate for an InSb wire such as in the experiments in reference~[\onlinecite{Mourik25052012}], namely: $\Delta= 0.25\text{m}e\text{V}$, $\alpha=0.2e\text{V} \textup{\AA} $, $g=50$, $m=0.013m_0$, where $m_0$ is the electron mass.
For completeness, we also tested the model for the system parameters from the Weizmann experiment~[\onlinecite{MotyExpt1}], but all figures were plotted with the parameters defined here.
%
%

\section{Spectrum and Parity}
\label{sec:SpectrumParity}

%
%
\begin{figure}[htbp]
	\centering
		\begin{subfigure}[b]{.22\textwidth}
		\centering	
		\includegraphics[width=\linewidth]{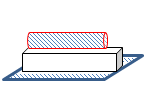}
		\caption{ }
		\label{fig:SetupNOSC}
	\end{subfigure}
	\begin{subfigure}[b]{.22\textwidth}
		\centering	
		\includegraphics[width=\linewidth]{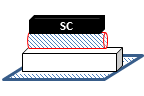}
		\caption{ }
		\label{fig:SetupSC}
	\end{subfigure}
	\begin{subfigure}[b]{.22\textwidth}
		\centering	
		\includegraphics[width=\linewidth]{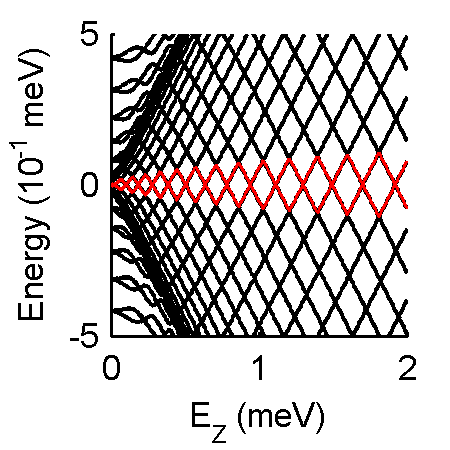}
		\caption{ }
		\label{fig:SpectrumNOSC}
	\end{subfigure}
	\begin{subfigure}[b]{.22\textwidth}
		\centering	
		\includegraphics[width=\linewidth]{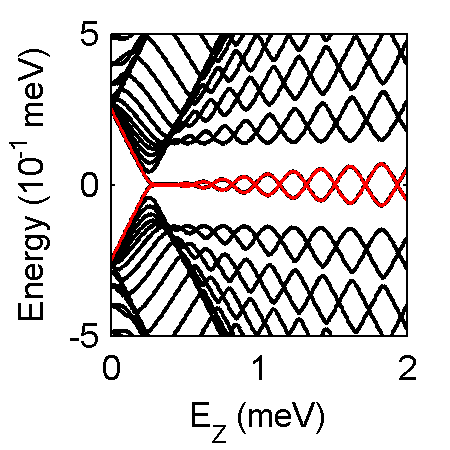}
		\caption{ }
		\label{fig:SpectrumSC}
	\end{subfigure}
	\caption{ \raggedright{(a)/(b) Schematic of geometry.  Semi-conducting nanowire sits on an insulating substrate (white), above a global backgate. Superconductor (in black) present in (b) but not (a). (c)/(d) Quasiparticle energy spectrum as function of Zeeman field $E_{\rm Z}$ for setups (a)/(b), with length $L=2\mu \rm{m}$, and other parameters $\mu=0$, $\alpha=0.2eV \AA$, $m^*=0.013m_0$, as defined in the text. Levels closest to zero are marked in red. In (c) we see discrete states from confinement. In (d), once in the topological regime $E_{\rm Z}>\sqrt{\mu^2+\Delta^2}$, we see the mid-gap degenerate Majorana states, which then split and oscillate (pair potential $\Delta=0.25 \rm{m}eV$).}}
	\label{fig:SetupSpectrum}
\end{figure}


\begin{figure}[htbp]
	\centering
		\begin{subfigure}[b]{.22\textwidth}
		\centering	
		\includegraphics[width=\linewidth]{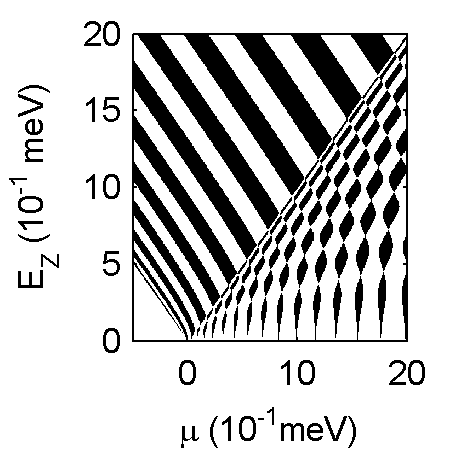}  
		\caption{ }
		\label{fig:ParityNOSC}
	\end{subfigure}
	\begin{subfigure}[b]{.22\textwidth}
		\centering	
		\includegraphics[width=\linewidth]{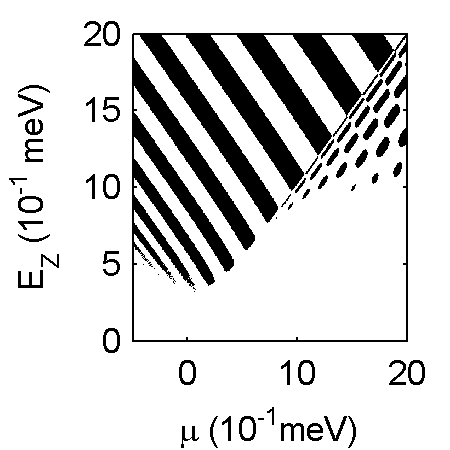} 
		\caption{ }
		\label{fig:ParitySC}
	\end{subfigure}
		\begin{subfigure}[b]{.22\textwidth}
		\centering	
		\includegraphics[width=\linewidth]{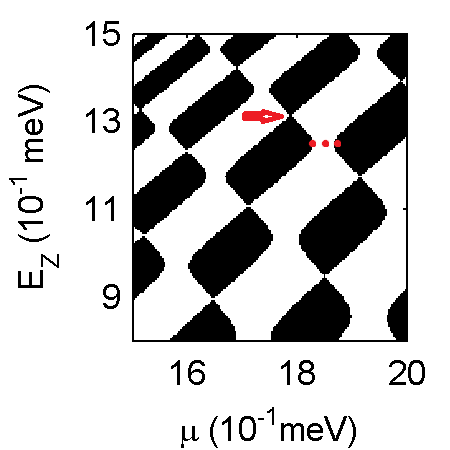}  %
		\caption{ }
		\label{fig:ZOOMParityNOSC}
	\end{subfigure}
	\begin{subfigure}[b]{.22\textwidth}
		\centering	
		\includegraphics[width=\linewidth]{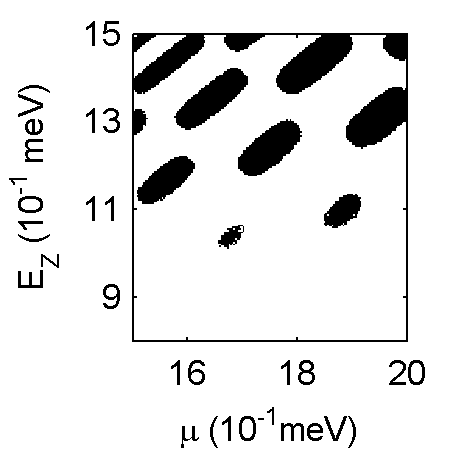}  
		\caption{ }
		\label{fig:ZOOMParitySC}
	\end{subfigure}
%
	\caption{\raggedright{Electron-number parity of the system, for (a) wire without a superconductor, and (b) wire proximity coupled to a SC. (c)/(d) Zoom on upper-right part of (a)/(b). Arrow highlights a double degeneracy. Dotted line marks an avoided crossing. System parameters: $g=50$, $\alpha=0.8eV\textup{\AA}$, $L=2\mu \rm{m}$, $\Delta=0$ in (a)/(c), and $\Delta=0.25 \rm{m}eV$ in (b)/(d).} }
	\label{fig:Parity}
\end{figure}
We begin by examining the case of a wire without a superconductor, to gain intuition.
The system we consider consists of a semiconducting nanowire with large Rashba spin orbit (SO) coupling sitting on an insulating substrate with a global back-gate below, as shown in Fig.~\ref{fig:SetupNOSC}.
The substrate is required to break inversion symmetry for Rashba SO, and the back-gate allows for control of the chemical potential.
Notice that we have a wire sitting on an insulator with no other contacts, not to be confused with the case of a wire connected to a metal whose superconducting gap is reduced, say by a magnetic field.
If we think of $\Delta$ in the wire as being dependent on $\Delta_{SC}$ of a superconductor and the tunneling between the wire and the superconductor, this is equivalent to taking the tunneling to zero while keeping $\Delta_{SC}$ fixed.
We nonetheless refer to this case as $\Delta=0$.
Fig.~\ref{fig:SpectrumNOSC} shows the spectrum for a wire without a superconductor as a function of magnetic field, calculated from Eq.~\ref{eq:Hamiltonian}.
The spectrum is not gapped, and the discrete states crossing the Fermi level are due to the finite length of the isolated wire.

The number-parity of the wire is plotted as a function of $\mu$ and $B$, in Fig.~\ref{fig:ParityNOSC}, with black regions corresponding to odd-parity states.
In this case, the parity is calculated directly from the electron number in the wire, since it is well defined without superconductivity.
The boundaries between regions of constant parity correspond to energy states crossing the Fermi level, as in Fig.~\ref{fig:SpectrumNOSC}.
These boundaries are the locus of points at which the system is compressible, and the charge changes discretely across these points as we fill each newly available state.
We therefore refer to these parity plots as charging diagrams.

At $B=0$, we see degenerate Kramer's pairs of opposite spin states, and thus no odd parity region.
As we increase $B$ these states are spin-split, and at high enough fields, all states at a given $\mu$ are spin-polarized.
For intermediate $B$ --- i.e. $0<E_{\rm Z}<\mu$ --- we see figure-eight patterns in the charging diagram where states avoid each other at some values of $\mu$ and $E_{\text{Z}}$ and cross at others.
These can be seen more clearly in Fig.~\ref{fig:ZOOMParityNOSC}, which focuses on a region of Fig.~\ref{fig:ParityNOSC}.
The avoided crossings (red dashed line in Fig.~\ref{fig:ZOOMParityNOSC}) are due to spin-orbit coupling, and the degenerate points (red arrow in Fig.~\ref{fig:ZOOMParityNOSC}) occur because the Rashba spin-orbit interaction only mixes opposite-spin states between wavefunctions with different spatial parity.

To see this, note that the Hamiltonian is a system of linear equations which mix the two spin species.
Since, for $\Delta=0$, $H_{BdG}$ commutes with $P\sigma_z$, where $P$ is the spatial inversion operator, it follows that states with different eigenvalues of $P\sigma_z$ have vanishing matrix elements.
Equivalently, by writing the wave function for each spin expanded in Fourier modes, it is obvious that mixing of opposite spins occurs only when the Fourier modes have opposite spatial parity.
We calculate the spacing between avoided levels $n$ and $m$, where level $n$ is the $n^{\text{th}}$ Kramer pair counting from $\mu=0$, in a wire of length $L$.
We find
\begin{equation}
\Delta E_{n,m} = \left| \frac{\alpha n m}{L(n^2-m^2)} \right|,
\end{equation}
where $B$ is implicitly included in the equation since larger $|n-m|$ means states only approach each other at higher $B$.
The avoided crossings can be used as another means to extract the value of $\alpha$, the spin-orbit strength.
Note that the spin-polarised states at high-$B$ and the avoided-crossing figure-eights are the only two distinct regimes in this $\Delta=0$ case.

Although it is not shown in the figures, we may also consider rotating the magnetic field from along the length of the wire, to the spin-orbit direction.
When the applied $B$ is parallel to the spin-orbit field, it is qualitatively equivalent to setting $\alpha=0$, although $\alpha$ does provide a quantitative shift to the result.
Indeed, the Zeeman split states no longer avoid each other, and just evolve linearly with $B$.
We note that an experiment in which the magnetic field is rotated until the avoided crossings completely disappear provides a clear measurement of the spin-orbit direction.
As discussed below, this can also be done for the wire-SC hybrid systems.

With these insights from the non-superconducting case, we consider a wire proximitised by a SC.
This is similar to the above setup, although the wire is now coupled to a large superconductor, which we treat as a bath as in Fig.~\ref{fig:SetupSC}.
We assume that the wire and superconductor are in thermal equilibrium, such that fermion parity can change on the time-scales of the experiment.
In Fig.~\ref{fig:SpectrumSC}, we show the spectrum for this case, under the same conditions as the non-superconducting case.
At low $B$, the system is gapped, and as a function of magnetic field the crossover from a non-topological state to a topological state is clear at $E_{\rm Z} = \sqrt{ \Delta^2+\mu^2}$ where the gap closes.
Within the topological regime, we see the two mid-gap states oscillating, with energy crossings that correspond to parity changes of the wire.
The splitting depends exponentially on the length of the wire, and goes approximately as $\exp{(-L/\xi)}\cos(k_F L)$, where $\xi$ is the superconducting coherence length in the wire, and $k_F$ the Fermi wave vector.
Note that the splitting increases with increasing Zeeman field, since $\xi$ increases with $B$~[\onlinecite{smokinggun, MotyExpt1}].

We calculate the number parity in the wire with induced superconductivity, and plot it as a function of $\mu$ and $B$, in Fig.~\ref{fig:ParitySC}.
The method used to calculate the parity is discussed in the Appendix.
The theoretical boundary between the topological and non-topological regimes corresponds to the curve
\begin{align}
E_{\text{Z}}=E_c \equiv \sqrt{\mu^2 + \Delta^2},
\end{align}
and in the limit $L \rightarrow \infty$, the parity is constant below this curve.
Below this boundary, the number of particles fluctuates as Cooper pairs are interchanged with the superconductor, but there are no changes in the parity.
Within the topological regime, $E_{\text{Z}}>E_c$, the mid-gap states have net spin polarisation and evolve linearly in the $\mu$-$B$ plane, similar to the wire without a superconductor.
Between parity flips, the density varies continuously.
Comparing the parity flips in Fig.~\ref{fig:ParitySC} with the oscillations in the BdG spectrum -- Fig.~\ref{fig:SpectrumSC}  -- we see that the flips correspond precisely to the degeneracy points between the Majorana modes.
This confirms that the parity flips are a signature of the split Majorana states crossing the Fermi energy.

We note that in an experiment, by fitting the outermost parity-flip, corresponding to the topological boundary, to the hyperbola $E_{\rm Z}=\sqrt{\Delta^2+\mu^2}$ for small $B$ and $\mu$, the value of the induced pair potential $\Delta$ can be obtained.
This is an important system parameter, whose value has an important effect on interpretation of experiments.
Although $\Delta$ has been measured through transport measurements, independent confirmation is important, especially given the recent discussion of soft gaps due to leads~[\onlinecite{Stanescu}].
However, this fitting procedure can be difficult, since at large $B$ and $\mu$, the topological boundary is only weakly dependent on $\Delta$; we show an alternate way to extract $\Delta$ at the end of Sec.~\ref{sec:TotalCharge} below.

An interesting new parameter regime to examine is large $B$ and $\mu$, outside the topological region, i.e. $E_c > E_{\text{Z}} > \Delta $, the upper right side of Fig.~\ref{fig:ParitySC}, and enlarged in Fig.~\ref{fig:ZOOMParitySC}.
Here we see parity flips, but they evolve quite differently from those within the topological region.
In this non-topological regime, $B$ is so strong that the wire is almost gapless, and the changes in parity are discrete and due to the finite length of the wire.
At large enough $B$, this is true on both sides of the topological boundary.
The figure-eight like patterns from the $\Delta=0$ case are no longer present, as the double degeneracy points have now become avoided crossings.

For completeness, we may consider turning off the spin-orbit interaction, killing the mid-gap Majorana states.
Setting $\alpha=0$ in our model, one finds that the system has even parity for all $\mu$ and $E_{\text{Z}}< \Delta$.
When $E_{\text{Z}}>E_c$, the alternating parallel parity stripes we saw in the other cases are present.
For  $\Delta < E_{\text{Z}}<E_c$, one finds a checkerboard pattern of constant parities formed by the two spin-states evolving in opposite directions with $E_{\text{Z}}$.
Although this behaviour for $E_{\text{Z}}<E_c$ distinguishes the $\alpha=0$ case from the $\alpha > 0$ case, we stress that both show very similar behaviour when $E_{\text{Z}}>E_c$.

For a ``topological'' wire with $\Delta>0$ and $\alpha>0$, if the applied magnetic field is rotated so that it has a component along the direction of the spin-orbit-field, the system begins to behave as if it has no spin-orbit interaction ($\alpha=0$).
This is analogous to the avoided crossings disappearing when the field is rotated in the $\Delta=0$ case, as discussed above.
When this perpendicular magnetic field component becomes strong enough compared to the axial field, the charging diagram and the peak heights look like the $\alpha=0$ case.
The spin-orbit does not couple opposite spins, and just adds to the Zeeman field.
It therefore no longer makes sense to discuss a topological regime.
Rotating the field perpendicular to both the wire and the spin-orbit direction has no effect on the charging diagrams.

In all of the above discussion, the parity-transitions indicate that the system is compressible at these points in parameter space --- it is possible to add charge.
In particular, we can calculate the change in charge as we cross these boundaries.
So far, it seems that in the three cases we have examined -- the topological case, $\Delta=0$, and $\alpha=0$ -- there are spin-polarised parity flips in the regime $E_{\rm Z} > E_c$, each qualitatively indistinguishable from the other cases.
In order to identify split Majorana fermions, we need to distinguish between these three cases.
To do so, we take a closer look at the size of the discrete charge jump across these parity-boundaries.
This change in charge can be detected through compressibility measurements.

\section{Jumps in charge density}
\label{sec:Jumps}
In this section we show that while the Majorana wave functions are localised at the two ends of the wire, the jumps in the charge density are roughly uniform across the wire. 
We show that this happens because, roughly speaking, the difference in the charge density of the even and the odd ground states is given by $\delta \rho(x)=|u(x)|^2-|v(x)|^2$ with $2u(x)=u_R(x)+iu_R(L-x)$ and $2v(x)=u_R(x)-i u_R(L-x)$, where $u_{R/L}(x)$ is an exponentially decaying (real) function peaked at the right/left end of the wire. 
One therefore obtains that $\delta \rho(x) =-u_R(x) u_R(L-x)$ is roughly uniform as the two exponential factors cancel each other.
We derive here the expression for $\delta \rho(x)$ by calculating the full expression for the wave-functions $u(x)$ and $v(x)$.

Following the supplementary material of reference~[\onlinecite{smokinggun}], we note that in the bulk of the wire, there are generally eight linearly independent solutions of the BdG differential equations at the energy $E=0.$  
There are four solutions in which  the spinor $u=\left( u_{\up}, u_{\down}   \right)^T$ is pure real and in spinor notation, $v= u^*=u $, and four in which $u$ is pure imaginary, and $v=u^*= -u$.  
The two classes are labeled, respectively  by an index $\lambda = \pm 1$. 
The general solution for a fixed $\lambda$ can be written as
\begin{align}
u^\lambda(x)  = \sum_{n=1}^4 a_n e^{-z_n x} \rho_n ,
\end{align}
where $z_n$ are  roots of the quartic equation
\begin{equation}
\label{B2}
(\frac{z^2}{2m} + \mu^2)^2 - E_z^2 + (z \alpha - \lambda \Delta)^2 = 0 ,
\end{equation}
and $\rho_n$ are two component spinors, independent  of $x$, whose explicit forms are given in reference~[\onlinecite{smokinggun}].   

For $\lambda = -1$, if the system parameters are  in a topological superconductor phase,  the quartic equation will have two complex conjugate  solutions, denoted by $(z_1,z_2)= z_{\pm}$ which have positive real parts, one positive  real solution, denoted $z_3=w$, and one negative solution, which we denote $z_4=s$.  
The spinors $\rho_n$ may be chosen such that  both components of $\rho_3$ and $\rho_4$ are real, while $\rho_1 = \rho_2^*$.  
Then, to obtain a solution with pure imaginary $u_\lambda$ , we must choose $a_3, a_4$ to be pure imaginary, and $a_2 = - a_1^*$.

For $\lambda=1$,  the solutions of Eq (\ref{B2}) will be written as $z'_n = - z_n$, where $z_n$ are the solutions for $\lambda=-1$ and the corresponding spinors are given by $\rho'_n = \rho_n$.
In the non-topological phase, there will be two solutions with positive real parts and two with negative real parts for both choices of $\lambda$.

For a semi-infinite wire, defined in the region $0<x<\infty$, we impose boundary conditions that $u=0$ at $x=0$ and that $u \to 0 $ for $x \to \infty$.   
For the case $\lambda=-1$, the second requirement is satisfied if and only if we choose $a_4=0$.   
This leaves us three real parameters, $a_3$ and the real and imaginary parts of $a_1$.  %
As the boundary condition at $x=0$ imposes only two additional conditions on $u$, we can always find a nonzero choice of the coefficients $a_n$ to satisfy all requirements.  
This defines the wave function for a zero-energy Majorana mode localised near $x=0$.    

For the case $\lambda=1$, the requirements that the wave function decay for $x \to \infty$ means that three coefficients must be chosen equal to zero,  corresponding to $n=$1, 2, and 3, leaving only one coefficient to adjust.  
Clearly this will not allow us to satisfy the boundary condition at $x=0$.   
In the non-topological regime, there are two adjustable coefficients for either choice of $\lambda$, which means that one cannot find a nonzero solution of the equations in either case.

Returning to the topological case, and following~[\onlinecite{smokinggun}], we may write
\begin{align}
z_1 = -i k_F  + \kappa,
\end{align}
where $k_F$ and $\kappa$ are positive and $k_F$ reduces to the Fermi wave vector of the normal wire in the limit where the pairing potential $\Delta$ is small.  
The envelope of the Majorana wave function will decay exponentially for $x \to \infty$ with a decay length $\xi$, given by 
\begin{align}
 \xi^{-1} = \min (\kappa, w)
\end{align}

In the limit where $E_{\rm Z}$ tends to the critical value $E_c =\sqrt{ \Delta^2 + \mu^2}$ for the transition to the non-topological phase, so the energy gap vanishes at $k=0$, one finds that $w \to 0$, and hence $\xi=w^{-1}$.   
However, for magnetic fields such that $E_{\rm Z}$ is larger than a second value $E_2$, one finds $w > \kappa$, so that $\xi = \kappa^{-1}$.
In this regime, the large distance behavior of the Majorana wave function may be written
\begin{equation}
\label{eq:B1}
u(x) \sim  e^{- \kappa x} \sin(k_F x + \phi) ,
\end{equation}
where the phase shift $\phi$ will itself be small for large values of $E_{\rm Z}$.  
As was noted in~[\onlinecite{smokinggun}], the crossover field  $B_2$ is fairly close to the critical value $B_c$ for nanowires such as InSb.
%

In the case of a long but finite wire, we must replace the boundary condition at infinity by the condition that the two components of $u$ should vanish at $x=L$.  
For a finite wire, we no longer require  $a_4 = 0$.  
Since $a_4$ must be real, however, this gives us only one additional parameter to choose, and one cannot find a non-trivial zero-energy  solution for general values of the control parameters $\mu, B$ and $L$.  
On the other hand, zero-energy solutions could exist on discrete surfaces of co-dimension unity in the control parameter space.  

In the regime of parameters where Eq.~\ref{eq:B1} applies, for the semi-infinite system, we expect to find  these zero energy solutions on surfaces close to  the points where $L k_F (\mu, B) =  n \pi$, where $n$ is an integer.

If $L/\xi$ is large compared to unity, then the magnitude of $a_4$ necessary to satisfy the boundary conditions at $x=L$ will be of order $e^{- L (w+\kappa)}$.  
The  non-zero value of $a_4$ requires a correction to $a_1, a_2$ and $a_3$ in order to continue to satisfy the boundary conditions at $x=0$, but this correction is obviously small when $L$ is large.

 If a non-zero wavefunction $\psi^L (x)$  is obtained in this manner for $\lambda = -1$, it can be used to construct a zero energy Majorana operator, $\gamma^L$, which will have maximum weight at the left end of the wire,  and decrease exponentially for large $L$.  
We define  
\begin{align}
\gamma^{L} = \sum_{\sigma} \int \left[  u_{\sigma}^{L}(x) \Psi_{\sigma} (x) + v_{\sigma}^{L}(x) \Psi_{\sigma}^{\dagger}(x) \right]\text{d}x
\end{align}

In order to satisfy the requirement  $({\gamma^L})^2=1$, the wave function must be normalized so that 
\begin{align}
2 \sum_{\sigma} \int_0^L | u^L_{\sigma} (x) |^2 dx  = 1
\end{align}
In order to have a precise definition one must still introduce a convention with regard to the overall sign of the wave function. 
Here, we adopt the convention that the sign of $ -i u(x) $ should be positive for $x$ slightly greater than zero, for the spin component aligned with the applied magnetic field.  
For large magnetic field values, such that $\kappa < w$, this implies that  
\begin{align}
- i u^L_{\downarrow} (x ) \approx  C \kappa^{1/2} e^{- \kappa x} \sin (k_F x) ,
\end{align}
where $C$ is a constant of order unity.

Following a similar procedure for the case $\lambda = 1$, we can construct a Majorana operator $\gamma^R$ associated with the right end of the wire.  
Again there is an arbitrariness of an overall sign, however, we can fix the sign by choosing the wave function as
\begin{align}
u^R (L-x) = - i u^L (x) \, , \,\,\,\,\, v^R(L-x) =  i v^L (x).
\end{align}
It is easy to show that $u^R$  satisfies the necessary equations and boundary conditions for $\lambda=1$, and  that the corresponding wave function $\psi^{R}$ is orthogonal to $\psi^L$ under the BdG metric. 
This means that $\{ \gamma^R , \gamma^L\} = 0$. 

We may now form a BdG fermion annihilation operator $\Gamma = (\gamma^R + i \gamma^L)/2$,  with the corresponding BdG wave function $\psi =( \psi^R + i \psi^L)/2 $.    
Although the charge density is zero for all $x$ in the Majorana states $\psi^R$ or $\psi^L$, the charge density associated with the wave function  $\psi$ is given by 
\begin{align}
\left\langle \rho(x)\right\rangle_\psi \,  & = \frac{ |v^R(x) + i v^L(x)|^2 -|u^R(x) +  i u^L(x)|^2  }{4}  \nonumber \\ 
&= - u^R(x) \, u^R (L-x) ,
\end{align}
which is generally not zero. 
Specifically, $ \left\langle \rho(x) \right\rangle_\psi $ is the difference in charge density when the state $\psi$ changes from unoccupied to occupied.

Let  $L$ and $B$ be fixed at specified values, and let $\mu_0$ be a value of the chemical potential $\mu$ for which there exist zero energy states for the given $L$ and $B$.  
Let us now consider a chemical potential $\mu =\mu_0 + \delta \mu$, where $|\delta \mu|$ is small.  
The system Hamiltonian will therefore be modified by the addition of a term $- \delta \mu \int \rho (x) dx.$
Then the wavefunction $\psi$ constructed above is no longer an exact solution of the BdG equations.
To lowest order in $\delta \mu$, however, it remains a solution of the BdG equations, and the energy of the state is given by first order perturbation theory as
$E_\psi =  - \delta \mu \, \delta N_\psi$, where 
\begin{align}
\delta N_\psi = - \int_0^L \, u^R(x) \, u^R (L-x) \, dx ,
\end{align}
which we may interpret as the net number of electrons associated with the  zero-energy level $\psi$.  

For $\delta \mu \neq 0$, the ground state of the system will  have  $\psi$  occupied  if and only if $E_\psi  < 0$.    
The relation  $E_\psi =  - \delta \mu \, \delta N_\psi$  implies that if $\delta \mu $ changes from a value slightly smaller than zero to a value slightly larger than zero, the total electron charge will always jump by a positive amount, given by the absolute value $|\delta N|$.  
It also follows that the value of the jump is given by the slope of the energy curve for $\delta \mu \to 0^+$:
\begin{align}
 \delta N \equiv |\delta N_\psi | = \lim_{\mu \to \mu_0^+} \frac { \partial E_\psi} {\partial \mu} 
\end{align}

In the case of large $B$ and $\kappa L > 1$, one finds for the zero energy state at $\mu=\mu_0$:
\begin{align}
\left\langle \rho(x) \right\rangle_\psi \, \sim \kappa e^{- \kappa L} \sin^2 (k_F x)  (-1)^{n+1}
\label{eq:MajoranaChargeDensity}
\end{align}
where $n= k_F L / \pi$.
The net charge associated with the state is  given by $\delta N_\psi  \approx  (-1)^{n+1} \kappa L e^{- \kappa L}$ .
%

\section{Numerical Calculations of Total Charge}
\label{sec:TotalCharge}
%
\begin{figure}[!thpb]
\centering
\includegraphics[]{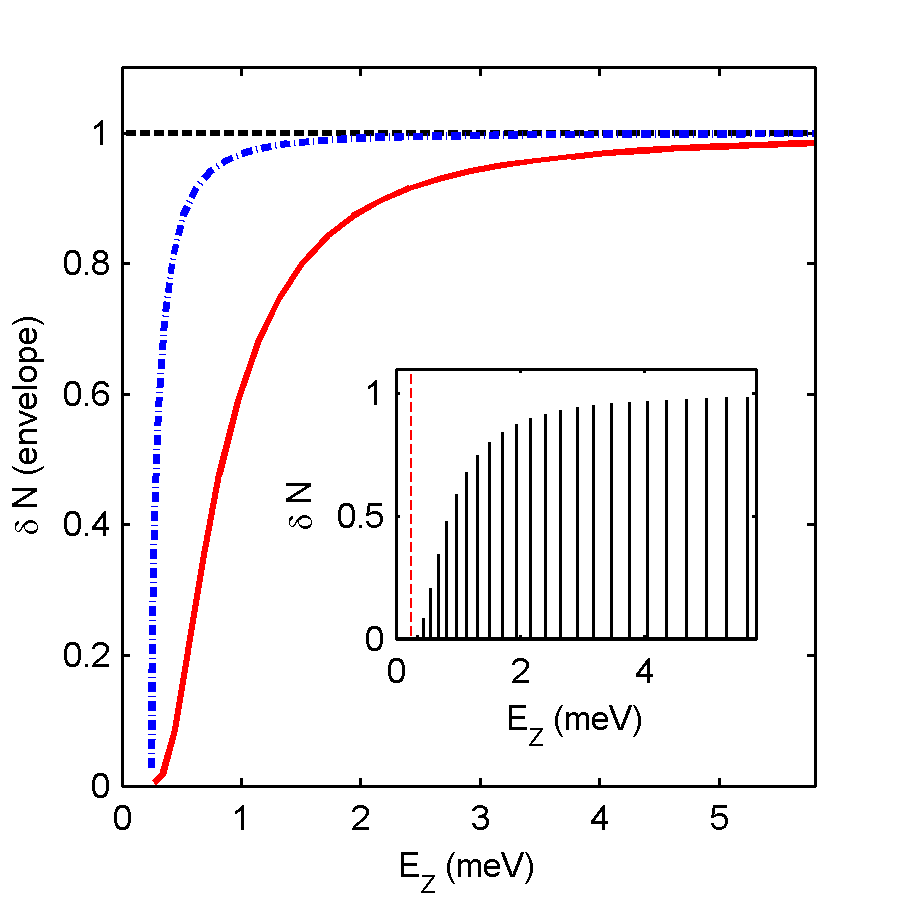}  %
\caption{\raggedright{Jumps in electron number in the wire, $\delta N$, as a function of $E_{\rm{Z}}$ for $\mu=0$. Inset: peaks correponding to the zeros of the spectrum in the topological regime. Red dashed line is at $E_{\rm{Z}}=\Delta$. Main figure: trace of the maximum of the peaks with interpolation. Solid red - $\alpha>0$, $\Delta>0$;  blue dash-dot - $\alpha=0$, $\Delta>0$ ;   black dashed - $\alpha=0$, $\Delta=0$.}}
\label{fig:PeakHeights}
\end{figure}
\noindent

We now calculate $\delta N$ numerically, using Eq.~\ref{eq:ChargeDensity}, and examine its dependence on the applied magnetic field.
The features described in Sec.~\ref{sec:SpectrumParity} can be traced to the cosine term in the splitting between the Majorana states.
The magnitude of the change in charge allows us to probe the $e^{-\kappa L}$ factor of Eq.~\ref{eq:MajoranaChargeDensity}, which also enters the energy splitting.
From this information, we can distinguish between topological charging events and non-topological states.

In particular, consider a plot of $\delta N$ in the wire at $\mu=0$ as a function of $E_{\rm{Z}}$, as shown in the inset in Fig.~\ref{fig:PeakHeights}.
The height of the peaks shows the magnitude of change in total charge in the wire.
The positions of these peaks correspond to the parity flips at $\mu=0$ as seen in Fig.~\ref{fig:ParitySC}.
At large $B$, the split-Majorana states saturate to one, since at high $B$ the splitting becomes comparable to $\Delta$, and the peaks represent discrete single-particle states.
The main part of Fig.~\ref{fig:PeakHeights} traces and interpolates between the maxima of these peaks for the various parameter regimes discussed, all at $\mu=0$.
For a wire without an induced superconducting gap -- regardless of the presence of spin-orbit interaction --  the peak height is constant and peaks are visible all the way down to $B=0$ (black dashed line in Fig.~\ref{fig:PeakHeights}).
This is as expected for a system without a gap, in which every charging event corresponds to the addition of an electron.
For a system with finite induced $\Delta$ and $\alpha=0$, we find that there are no peaks visible for $E_{\rm Z}<\Delta$, as expected when the system is gapped and there are no mid-gap states.
At large $ E_{\rm Z} >> \Delta $, the discrete charging events correspond to the addition of electrons, and $\delta N=1$.
When $E_{\rm Z} \sim \Delta$, the peaks correspond to a change in charge of less than one electron.

When $\alpha>0$, the topological case, peaks begin to appear at $E_{\text{Z}} \sim E_2 > \Delta$, and the magnitude has the form of the prefactor in Eq.~\ref{eq:MajoranaChargeDensity}.
Since $\kappa \sim 1/E_{\text{Z}}$ for $E_{\text{Z}}>E_2$ (see~[\onlinecite{smokinggun,halperin2012adiabatic}]), the dependence on the Zeeman field, solid red line in Fig.~\ref{fig:PeakHeights}, is roughly $\exp(-L/E_{\text{Z}})/E_{\text{Z}}$.
The shape of the curve is closely linked to the overlap -- and splitting -- of the Majorana modes.
The difference between the different traces of the amplitude height is a useful tool to distinguish between the oscillations in an experiment.
The calculations shown here are done at T=0.
At finite temperatures, we expect the discrete jumps to be smeared.
Since, at $\mu=0$, $E_2 \sim E_c=\Delta$, an experiment can extract a value for $\Delta$ from a plot such as Fig.~\ref{fig:PeakHeights}.

\section{Numerical Calculations of the Charge Distribution}
\label{sec:ChargeDistribution}
%
\begin{figure}[htbp]
	\centering
		\includegraphics{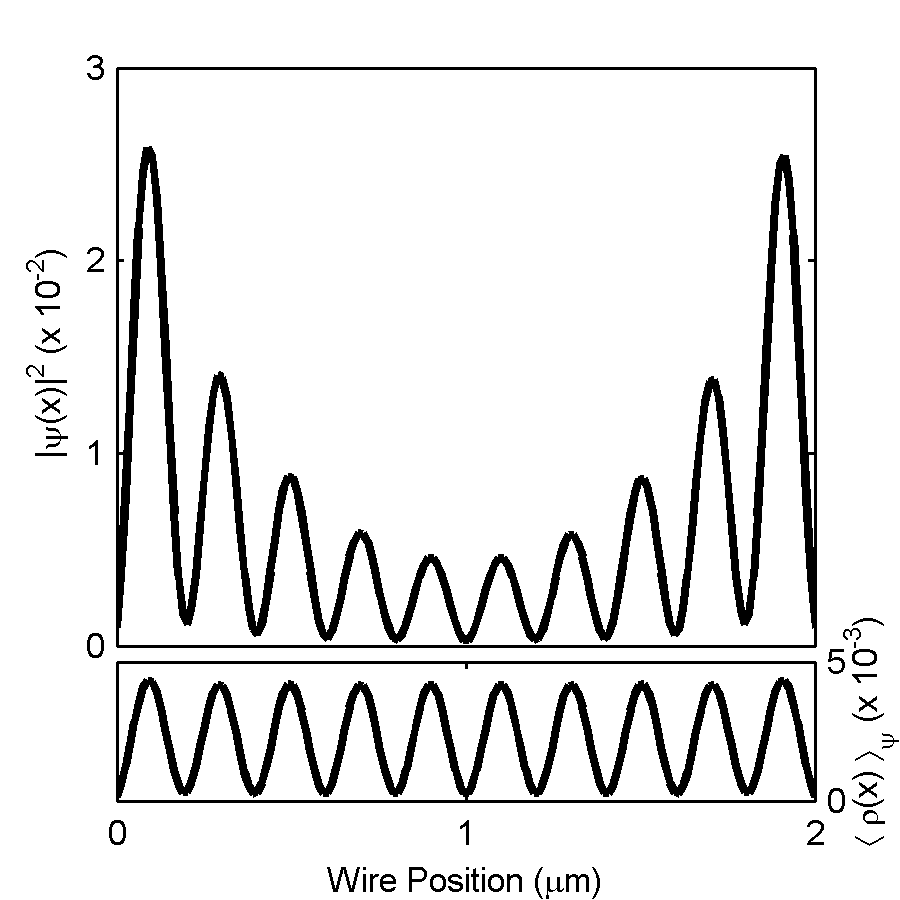}  
	\caption{ \raggedright{Top: Intensity  $|\psi(x)|^2 = |u|^2 + |v|^2$ of the wavefunction for a Majorana-pair state whose energy crosses zero at a degeneracy point in the spectrum ($E_{\rm{Z}} \sim 0.69 \rm{m}eV$). As expected, the wavefunction is concentrated at edges, and decays toward the centre. Bottom: The change in charge density, $ \left\langle \rho(x) \right\rangle_\psi = |v|^2 - |u|^2$, when this state becomes occupied. The charge is small but not zero, and is spread uniformly along the wire length.}}
	\label{fig:chargedist}
\end{figure}

%
%
%
Having established that the charge in the wire changes whenever the split Majorana states are degenerate, we now examine how the charge is distributed along the wire.
Since the discrete charging events within the topological regime correspond to the mid-gap state (as in Eq.~\ref{eq:MajoranaChargeDensity}), we examine the wave-function and charge of that state alone.
In the BdG basis chosen above, we calculate the amplitude $|\psi(x)|^2 = |u|^2+|v|^2$ and the charge $\left\langle \rho(x) \right\rangle_\psi = |u|^2-|v|^2$ as a function of position along the wire.
As in previous works ([\onlinecite{PhysRevLett.109.266402,halperin2012adiabatic}]), we see that the Majorana state is concentrated on the edges, as shown in the upper panel of Fig.~\ref{fig:chargedist}.
However, the charge corresponding to this state -- when the wires overlap -- is spread out along the wire (bottom of Fig.~\ref{fig:chargedist}).
Similar results may be seen in Fig.~6 of reference~[\onlinecite{PhysRevB.86.224511}].
Fig.~\ref{fig:chargedist} is calculated at a degeneracy point in the spectrum at $\mu=0$ and $E_{\rm{Z}} \sim 0.69 \rm{m}eV$, well inside the topological regime, with $E_{\text{Z}}>E_2$.
Near this point, the splitting decays and oscillates, as discussed in~[\onlinecite{smokinggun}].
The charge is distributed sinusoidally across the wire, implying that a measurement of the charge does not need to be done near the end of the wire.
Furthermore, the fact that the charge is distributed along the whole wire can be used to distinguish between the various explanations of the zero-bias conductance peak seen in transport measurement, since any non-topological causes should not have a uniform charge distribution.

\section{Jumps in Spin Density}
\label{sec:Spin}
Jumps in parity will generally be accompanied by jumps in the electron spin density as well as the charge density.
The jump in total spin $\delta<\vec{S}>$ will be given by 
\begin{align}
\left\langle \delta \vec{S} \right\rangle = \pm \hat{b} \frac{\partial E_{\psi}}{\partial E_{\text{Z}}},
\end{align}
where $\hat{b}$ is a unit vector in the direction of $\vec{B}$ and the sign in front is given by the sign of $\delta N_{\psi}$. 
The ratio between $|\left\langle \delta \vec{S} \right\rangle|$ and $\delta N$ is fixed by the Clausius-Clapyron relation which states $\delta \left\langle \vec{S}\right\rangle / \delta N$ is equal to the slope of $d\mu / dE_{\text{Z}}$ of the locus of parity jumps in the $\mu$-$E_{\text{Z}}$ plane.
The discontinuity in spin density should be uniformly spread along the length of the wire in a manner similar to the jumps in charge density.

\section{Effects of Electron-Electron Interactions}
\label{sec:Interactions}
Although our calculations, so far, have been based on a model with non-interacting electrons, we present here a brief discussion of the modifications one might expect due to Coulomb interactions in a real system. 

In general, one would expect that electron-electron interactions will renormalise parameters of the model, so that, e.g., $\mu$ and $\Delta$ may depend in a non-trivial way on the applied magnetic field and on the voltage applied to a nearby gate.  
However, we expect that a renormalised single-particle description will remain valid at low energies.  
Therefore, we expect that interactions will change the positions in the magnetic field and gate voltage where jumps in the number parity occur, but will not have a major effect on the size of the associated jumps in the charge of the nanowire, provided that the size is computed with a decay length $\xi$ appropriate to the renormalised values of $\Delta$ and the Fermi velocity.  
Our argument that the charge jump due to change in occupancy of a zero-energy Majorana pair should be roughly uniform along the length of the wire should be unchanged.  
At the same time, a parity jump due to a change in occupancy of, say, a localised impurity state, would produce a charge-density change in the nanowire that would remain at least partially localised in the vicinity of the impurity.   
We note that due to screening by the adjacent superconductor, the effective interaction between electrons on the nanowire will be relatively short-ranged.

Of course, screening by the superconductor will reduce the charge sensitivity of a nearby SET.  
However, we argue, using a simple model, that this effect should not be drastic.  
Therefore, we expect that SET measurements could be used to study the size of charge jumps in a real experiment, and could be used to distinguish a jump that is uniform along the wire from one that is concentrated at an impurity or at the ends of the nanowire.

Let us consider the voltage $V(y,z)$ measured at a point $(x,y,z)$, which is a distance $R=\sqrt{y^2+z^2}$ from the axis of the nanowire, at a position $x$ along the length of the wire, with $r_w<<R<<L$, where $r_w$ is the radius of the nanowire. 
The electrostatic potential $V$ at  the specified point should have the form
\begin{align}
\label{VR}
V(x,y,z) = \int_0^L dx'   \rho(x') K(x',x,y,z),
\end{align}
where $\rho$ is the charge density in the nanowire and the kernel $K$ depends on the detailed geometry. 
We expect that $V$ should be most sensitive to the charge density at points where $|x' - x| \leq R$, so as  a crude approximation we may write 
\begin{align}
\label{Vapprox}
V(x,y,z) \approx \tilde{\rho}(x) C(y,z)
\end{align}
where $\tilde{\rho}(x) $ is an average of the charge density over the region  $|x' - x| \leq R$ and $C(y,z) $
again depends on the geometry of the system.

We may now envision an experiment with, say, three SETs, localised at different positions $x$ but the same distance $R$ from the wire.
We may position one SET at the centre of the wire ($x=L/2$) and the other two near the ends, $x=x_0$, and $x=L-x_0$, where $x_0$ is larger than $R$ but smaller than the superconducting decay length $\xi$.
The prediction of our analysis,  combined  with the approximation (\ref{Vapprox}), is that a charging event due to a change in the occupation of a zero energy Majorana pair should cause a voltage jump with the same strength at all three detectors. 
By contrast, if the charging event were concentrated at the two wire ends in the same way as the Majorana wave function itself, one would expect the voltage signal to be larger at the two ends than at the central SET. 
If the charging event were associated with an impurity at an arbitrary point in the wire, the voltage signals would in general be different on all three SETs, and might vary randomly from one event to another.

More properly, one should not use the approximation (\ref{Vapprox}) but rather the non-local relation 
(\ref{VR}) to analyse the charge distribution in the nanowire.  
However, if the kernel $K$ is  known, either from a calculation or from experimental calibrations, it should be relatively easy to distinguish between the different charge distributions considered above.

In order to estimate the coefficient $C(y,z)$, we consider a simplified model.
We suppose that the superconductor is represented by a perfect conductor of radius $r_s$, parallel to the nanowire, with an axis displaced from that of the nanowire by a distance $D$ which is of the order of $r_w+r_s$. 
We assume that the point $x$ is far from the ends of the wire compared to $R$, so we may treat the wires as infinite.
Further, we approximate the nanowire as a uniform line charge with a fixed density $\tilde{\rho}$, located on the line $y=z=0$.

Under these assumptions, we expect an image line charge a distance $d$ above the infinite semiconducting wire, and we expect it to lie within the cylindrical SC.
For two wires (charge and image charge), we have the potential at a point $(x,y,z)$:
\begin{align}
V(y,z)&=\frac{\tilde{\rho}}{4 \pi \epsilon_0} \left[ \ln(y^2+z^2) - \ln(y^2+(z-d)^2) \right] + \nonumber \\
&~~+ \eta\tilde{\rho},
\end{align}
where $\eta$ is the value of the potential at infinity.
We want the potential to vanish on the surface of the SC.
Setting the potential to zero, we find that the potential vanishes on a circle, and by setting the radius to be $r_s$, we can solve for
\begin{align}
d=\frac{2r_w r_s+r_w^2}{r_s+r_w},
\end{align}
and
\begin{align}
\eta=-\frac{1}{2\pi\epsilon_0} \ln \left[  1+ \frac{r_w}{r_s} \right].
\label{eq:eta}
\end{align}
This gives
\begin{align}
C(y,z)&=\frac{1}{4 \pi \epsilon_0} \left[ \ln(y^2+z^2) - \ln(y^2+(z-d)^2) \right] - \nonumber \\
&~~-\frac{1}{2\pi\epsilon_0} \ln \left[  1+ \frac{r_w}{r_s} \right].
\end{align}

The analysis above may be extended to the case where the charge density on the nanowire has the form
\begin{align}
\rho (x) = \rho_q \cos qx ,
\end{align}
where the wave vector $q$ is assumed small compared to $1/r_w$.
In this case, the charge on the superconductor will not precisely cancel the charge on the nanowire, and there will be a component of the potential which depends logarithmically on $R$, in the region $r_w<R< 1/q$, while the potential falls  to zero  for $R \gg 1/q$.  More precisely, for $q \neq 0$,  one finds 
$V(x, y, z) = K_q \rho (x)$, with 
\begin{align}
K_q \approx  \eta \left[ 1 -  \frac{ \ln (R/r_w)}{\ln (qr_w)} \right]  
\end{align}
in the region $r_w < R <1/q$, where $\eta$ is the quantity given by Eq.~(\ref{eq:eta}).  
Thus, $K_q$ reduces to our previous result for $C(y,z)$, in the limit $q \to 0$, with  $R/r_w$ fixed but large.
For an infinite wire, the dependence of the kernel $K(x',x, y,z)$ on the separation $x'-x$  may be obtained by taking the Fourier transform of $K_q$.  
The logarithmic dependence of $K_q$ means that $K$ will not fall off very rapidly for $|x'-x| \gg R$.

For $R>>d$, taking $r_s=2r_w$ and $\tilde{\rho}= 0.1e/L$ where $e$ is the electron charge and $L=2\mu \rm{m}$, we find $V(R) \sim 60 \mu V$, which should be detectable with a SET.

So far, we have assumed implicitly that there is just a single
contributing mode in the nanowire.  
In the case of a multi-mode wire, any charge inhomogeneity due to a localised impurity state will be further screened by the additional modes in the wire, which will tend to spread the resulting charge more uniformly along the wire.   
This will reduce the differences in the voltages measured by SETs at different positions along the wire, but it should not affect the average voltage signal.  
The extra modes should not affect the signal induced by a spatially uniform charge jump, such as predicted due to the change in occupancy of a zero-energy Majorana pair.

\section{Experiment}
\label{sec:Experiment}
As we have argued above, charge jumps in the semiconductor nanowire
should be observable using a single electron transistor (SET) as a
sensitive charge detector[\onlinecite{PhysRevB.86.224511,Ilani1,Jens1}], assuming that the wire length
$L$ is not too much longer than the coherence length $\xi$. 
Furthermore, measurements at several positions – either through multiple or scanning SETs – can be done to confirm the uniform charge distribution.
We note that this measurement technique can be applied to other systems expected to have Majorana end states.
In particular, Majorana states in wires made from other materials, or created within 2D topological insulators (e.g. HgTe quantum wells~[\onlinecite{AmirHgTeTheory}] ), if realised, can hopefully be observed with an SET.
Furthermore, in these systems, multiple-band concerns might be alleviated.

An important experimental parameter which we hold fixed in our discussion is the wire length.
For the case of a long wire, the Majorana end states are present, but the splitting between the two states is exponentially suppressed, and therefore the number-parity oscillations are harder to observe.
Simultaneously however, a longer wire means smaller level spacing, and therefore more oscillations with respect to $B$ before the splitting reaches the size of the gap.
We thus conclude that there is an intermediate range ideal for experiments, where the exact length desired depends on the other system parameters.
For a non-topological wire -- $\alpha =0$ or $\Delta =0$ -- the level spacing decreases with wire length, until the system is compressible everywhere in the $\mu$-$B$ plane.
Our calculations are consistent with these expectations.

A related possible experiment is to measure the jumps in the spin of these wires.
The split Majorana states carry spin in addition to their electric charge.
This spin is considerably smaller than the spin of a single electron, and therefore very difficult to detect using available experimental techniques~[\onlinecite{Grinolds,Rugar2004}].
However, recent advances suggest that such measurements might not be so far off~[\onlinecite{ZELDOV,Rugar2013}].
With an extremely sensitive magnetometer, we can hope to pick out the oscillations in the magnetisation of the system as a function of $\mu$ and $B$, as discussed for the charge.

%
\section{Conclusion}
\label{sec:Conclusion}
In short topological wires, the predicted zero-energy Majorana end modes are split due to the significant overlap of their wavefunctions.
The split states carry charge, which can be detected in experiments.
Whereas the tunneling density of states measured in transport experiments is only an end effect, the charge of the split Majoranas is uniformly distributed along the wire.
Comparing both charge and tunneling experiments at the end and bulk of a wire can thus resolve remaining unanswered questions in the field.

%
%
\section{Acknowledgments}
\label{sec:Acknowledgments}
We thank A.~Akhmerov, E.~Berg, D.~Chowdhury, K.~Flensberg, G.~Gervais, S.~Hart, C.R.~Laumann, C.~Marcus, J.D.~Sau for useful discussions. 
GB supported by NSERC. 
Work was supported in part by a grant from the Microsoft Corporation and by the STC Center for Integrated Quantum Materials, NSF grant DMR-1231319. 
The work at WIS was supported by WIS-TAMU, ISF, Minerva and ERC (FP7/2007-2013) 340210 grants.
AY is also supported in part by DMR 1206016
%

%
%
\appendix
\section{Methods for Calculating Number Parity}
We discuss two equivalent numerical methods for calculating the number parity when $\Delta >0$. The plots in Fig.~\ref{fig:Parity} were actually obtained using the second method, but both methods were checked against each other.

The first  option is to follow the energy eigenvalues along a curve in the $\mu$-$B$ plane, starting at a point with $B=0$, and ending at the desired point $(\mu,B)$.
We know that the number parity must be even when $B=0$, and the number parity will flip when and only when an energy level crosses zero.
Therefore, the number parity at $(\mu,B)$ is equal to $(-1)^n$, where $n$ is the number of zero-energy crossings along the curve. 
Numerically, some care must be taken to correct for errors where two consecutive zeros are so close to each other that they appear as one,  resulting  in the wrong parity being recorded beyond the second of the close points.

As an alternative, we have used a new method, which to our knowledge has not been previously discussed in the literature. 
For a spinful system on a lattice with $N$ sites, write $H = \vec{\psi}^{\dagger} \mathcal{H}_{\text{BdG}} \vec{\psi} $, with $\vec{\psi} = (a_1, ..., a_{2N}, a_1^{\dagger}, ... , a_{2N}^{\dagger})^T$.
Then $\mathcal{H}_{\text{BdG}}=U \mathcal{D} U^{\dagger}$, where   $U=\left(  \begin{array}{c c}  u & v \\ v^* & u^*   \end{array} \right)$  is a unitary matrix and  $\mathcal{D}$  is a diagonal matrix  ordered so that the the first $2N$ elements are the positive energy eigenvalues.
We thus have:
\begin{align}
H &= \vec{\psi}^{\dagger} U \mathcal{D} U^{\dagger} \vec{\psi} \nonumber \\
&= \vec{\eta}^{\dagger} \mathcal{D} \vec{\eta},
\end{align}
with $ \vec{\eta} = (\eta_1,... ,\eta_{2N},  \eta_1^{\dagger},... ,\eta_{2N}^{\dagger})^T$.
We claim that the parity of the system is $P= (-1)^q$, where $q=\text{rank} (v)~\text{mod} 2$.
We have checked this numerically for $\Delta \geq 0$, and prove it for $\Delta=0$, along with a slightly different version of the claim for the case $\Delta>0$.
In particular, for $\Delta>0$, we will show  that $\det(v) \neq 0$ if and only if the system is in an even-parity state, subject to the following assumption, which we find compelling. 
Specifically, since the pairing term in the Hamiltonian does not conserve electron number, we assert that  a ground state with even number parity should contain some admixture of states with every possible even electron number between zero and $2N$, including the single basis state with $2N$ electrons present. 
Thus, we shall assume that if the ground state $ \left| G \right\rangle$ has even number parity, then
\begin{align}
\label{Ag}
\left\langle G \right |  a_1^{\dagger} \cdot \cdot \cdot a_{2N}^{\dagger}   \left| 0 \right\rangle \neq 0.
\end{align}

\begin{proof}[Proof for $\Delta=0$]

In the absence of a pairing potential an occupied eigenstate of $\mathcal{H}_\text{BdG}$ corresponds to a vanishing column in $u$ and a non-zero column in $v$, whereas for an unoccupied eigenstate the converse is true. This means that the number of occupied states $n_{\rm{occ}}$ simply equals the number of non-zero columns in $v$ which, since $U$ is unitary, are all linearly independent. One then has by definition that $n_{\rm{occ}}=\text{rank}(v)$, and in particular $q=\text{rank} (v)~\text{mod} 2$ as claimed.
\phantom\qedhere
\end{proof}
\begin{proof}[Proof for $\Delta>0$]

Consider the state 
\begin{align}
\label{Agp}
\left| G'  \right\rangle \equiv  \eta_{2N} \cdot \cdot \cdot \eta_1 \left| 0 \right\rangle.
\end{align}
Since any $\eta_i$ operating on this state annihilates it,  $ \left| G' \right\rangle$ must be proportional to the ground state, unless it is identically zero. That is,  $ \left| G' \right\rangle = C  \left| G \right\rangle$, for some constant $C$.   
Furthermore, if we transform the $\eta_i$'s in (\ref{Agp}) back to the electron operator basis, it is straightforward to show that the term $  a_1^{\dagger} \cdot \cdot \cdot a_{2N}^{\dagger}   \left| 0 \right\rangle$ occurs with a coefficient equal to 
 $\det(v)$.
According to our assumption (\ref{Ag}),  this term cannot have zero weight in the even-parity ground state, and therefore $\det(v) \neq 0$, which further implies that $v$ must have maximal rank.

Conversely, if $\det(v)=0$, then the system does not have  a component containing $2N$ electrons, so by our assertion, it cannot be an even-parity state. 
We thus have that $\det(v) \neq 0$ if and only if the  number parity of the ground state is  is even.

Although this result is sufficient for our purposes, if we make an assumption  analogous to (\ref{Ag})  for the case where the ground state has odd number parity,  namely that the $2N-1$ electron state must have non-zero weight in the ground state, we see that the pre-factor in the expansion from $(2N-1) $ $\eta_i$'s to the electron operator basis must be non-zero.
In particular, one can show that the pre-factor of the leading term is now a weighted  sum of the first minors of the matrix $v$, and by the same argument as above it cannot vanish.
A vanishing determinant with a non-vanishing first minor implies that a matrix has rank one less than its maximal rank, and so for the odd case, $\text{rank}(v)=2N-1$.
Since $v$ has size $2N \times 2N$, we see that the ground state number parity is given by $P = (-1)^q$, where   $q= \text{rank} (v)~\text{mod}  2 $, as claimed.  
\end{proof}



\bibliography{Dots_Biblio}

\end{document}